\documentclass[aps,prl,showpacs,floatfix,twocolumn,letterpaper]{revtex4-1}
\usepackage[pdftex]{graphicx}
\usepackage{amsmath}
\usepackage{hyperref}
\usepackage{latexsym}
\usepackage{color}
\usepackage{amssymb}
\usepackage{amsmath}

\begin{document}

\title{Controlling the scattering length of ultracold dipolar molecules}

\author{Lucas Lassabli{\`e}re}
\affiliation{
Laboratoire Aim\'{e} Cotton, CNRS, Universit\'{e} Paris-Sud, ENS Paris-Saclay, Universit\'{e} Paris-Saclay, 91405 Orsay, France 
}

\author{Goulven Qu{\'e}m{\'e}ner}
\affiliation{
Laboratoire Aim\'{e} Cotton, CNRS, Universit\'{e} Paris-Sud, ENS Paris-Saclay, Universit\'{e} Paris-Saclay, 91405 Orsay, France 
}

\date{\today}

\begin{abstract}
By applying a circularly polarized and slightly blue-detuned microwave field  with respect to the first excited
rotational state of a dipolar molecule, one can engineer a long-range, shallow potential well
in the entrance channel of the two colliding partners.  
As the applied microwave ac-field is increased, the 
long-range well becomes deeper and can support a certain numbers of bound states,
which in turn bring the value of the molecule-molecule scattering length 
from a large negative value to a large positive one.
We adopt an adimensional approach where the molecules are described
by a rescaled rotational constant 
$\tilde{B} = B/s_{E_3} $ where $s_{E_3}$ is a characteristic dipolar energy.
We found that molecules with $\tilde{B} > 10^8$
are immune to any quenching losses when a sufficient ac-field is applied, the ratio 
elastic to quenching processes can reach values above $10^3$, and that
the value and sign of the scattering length can be tuned.
The ability to control the molecular scattering length  
opens the door for a rich, strongly correlated, 
many-body physics for ultracold 
molecules, similar than that for ultracold atoms. 
\end{abstract}

\maketitle

Controlling the scattering length $a$ between ultracold particles 
is at the center of most modern ultracold gases experiments.
The scattering length corresponds to an effective parameter that characterizes
the range of the particles interaction at ultra-low energy.
The value and the sign of the scattering length control 
the interactions strength and stability of such gases
\cite{Dalfovo_RMP_71_463_1999,Leggett_RMP_73_307_2001}. 
A weakly interacting gas is defined when
the scattering length $a$ is much smaller 
than the mean relative distance $\bar{d}$ between the particles, 
$|a| / \bar{d} \ll 1$.
In contrast, a strongly interacting gas is defined when $|a| / \bar{d} \gg 1$ and leads to
a strong correlated state of matter \cite{Bloch_RMP_80_885_2008,Bloch_NP_8_267_2012}.
At the unitary limit, the scattering length diverges to an infinite value, positive or negative. 
With fermionic particles,
the strongly interacting regime represents a cross over between the BEC to the BCS weakly interacting regimes, 
when the very large scattering length changes sign from positive to negative
\cite{Regal_PRL_92_040403_2004,Bartenstein_PRL_92_120401_2004,Zwierlein_PRL_92_120403_2004,Bourdel_PRL_93_050401_2004}.
With bosonic particles,
few-body physics becomes strongly universal
\cite{Braaten_PRep_428_259_2006}
as underlined by the Efimov effect \cite{Efimov_PLB_33_563_1970,
Kraemer_N_440_315_2006}. 
Finally, controlling the scattering length of particles in optical lattices 
is very important to engineer tunable many-body Hamiltonians, 
to simulate untractable systems of condensed matter
\cite{Jaksch_AP_315_52_2005,Baranov_CR_112_5012_2012}.

In experiments of ultracold atoms, the control of the scattering length is usually possible in the vicinity 
of a Fano-Feshbach resonance \cite{Feshbach_AP_5_357_1958,Fano_PR_124_1866_1961}
when a magnetic field is tuned to an appropriate value
\cite{Tiesinga_PRA_47_4114_1993,Courteille_PRL_81_69_1998,Inouye_N_392_151_1998}. 
However, in experiments of ultracold molecules, for example ultracold alkali dipolar molecules  \cite{Ni_S_322_231_2008,Aikawa_PRL_105_203001_2010,Takekoshi_PRL_113_205301_2014,
Molony_PRL_113_255301_2014,Park_PRL_114_205302_2015,Guo_PRL_116_205303_2016,
Rvachov_PRL_119_143001_2017},
finding a well resolved, isolated Fano-Feshbach resonance is a difficult task
because of the very high density of states of tetramer bound states 
in the vicinity of the low-energy collisional threshold \cite{Mayle_PRA_85_062712_2012,Mayle_PRA_87_012709_2013}.
Even worse, this very high density of states yields long-lived tetramer complexes 
explaining losses in elastic collisions of non-reactive molecules \cite{Ye_SA_4_eaaq0083_2018}.
Therefore, the ability to tune the scattering length seems compromised for molecules.

In this paper, we show how we can control the molecule-molecule scattering length, 
which in this case becomes a complex quantity
$a = a_\mathrm{re} - i \, a_\mathrm{im} $ 
with $a_\mathrm{im} \ge 0$
\cite{Balakrishnan_CPL_280_5_1997,Hutson_NJP_9_152_2007}.
By applying a microwave field slightly blue-detuned with respect to the first excited
rotational state of the molecule, one can: 

\indent (i) bring the ratio good to bad collisions $\gamma = {\beta}_\mathrm{el}/{\beta}_\mathrm{qu}$ (elastic over quenching rate coefficient) 
to high values such that evaporative cooling techniques can be successful,

\indent (ii) suppress the imaginary part $a_{im} \to 0$ and shield the molecules against losses,

\indent (iii) tune the real part to small or large values, positive or negative
and control the interaction strength of an ultracold molecular gas.

\begin{figure}[t]
\includegraphics[height=30mm,width=50mm]{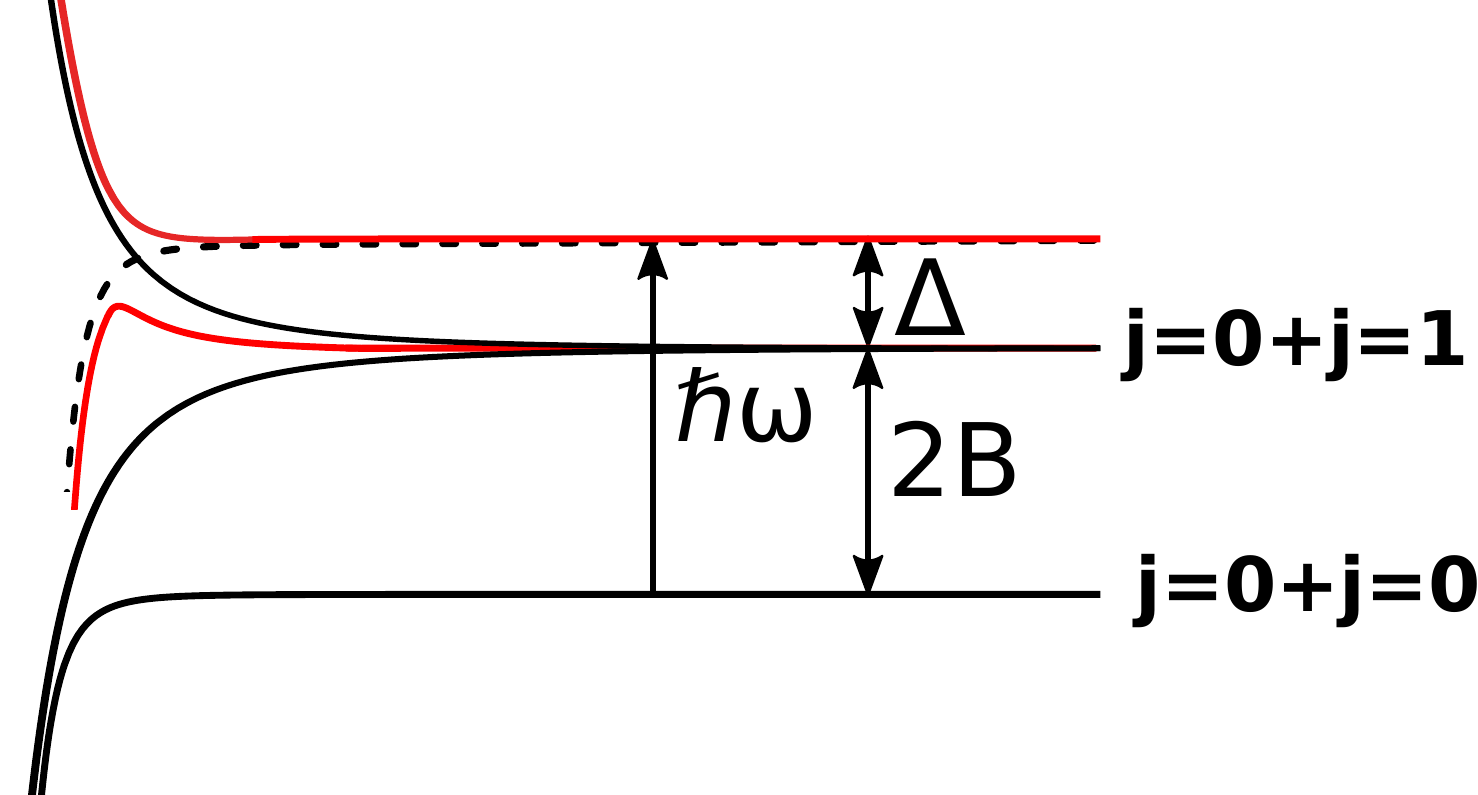}
 \caption{(Color online). Schematic process of a collisional shielding 
 of ground state rotational molecules $j=0$, using
 a blue-detuned, circularly polarized microwave field. $\Delta >0$
 is the detuning between the energy of the microwave field $\hbar \omega$
 and $2 B$, the energy level of the first excited rotational state $j=1$ of a molecule.
 The dipole-dipole interaction creates an effective repulsive adiabatic curve (plotted in red),
 preventing the molecules to approach at short-range.}
 \label{fig-scheme}
\end{figure}

\noindent By tuning in this way the scattering length at will, one can access 
with ultracold molecules the same rich and flexible, strongly correlated many-body physics of ultracold atoms as mentioned above. 
The basis of the method comes from the idea of optical shielding 
\cite{Suominen_PRA_51_1446_1995,Suominen_JPBAMOP_29_5981_1996,Napolitano_PRA_55_1191_1997,Weiner_RMP_71_1_1999}.
The schematic process is illustrated in Fig.~\ref{fig-scheme}. 
Instead of having an optical transition
slightly blue-detuned between an $s$ to a $p$ electronic state of an atom, 
one has a microwave transition of energy $\hbar \omega$ \cite{Micheli_PRA_76_043604_2007,Gorshkov_PRL_101_073201_2008,Karman_arXiv_1806_3608_2018}
between a $j=0$ to a $j=1$ rotational state of energy $2 B$, where $B$ is the
rotational constant of the molecule. 
The detuning is given by $\Delta = \hbar \omega - 2B > 0$. 
The advantage of a microwave shielding 
lies in the fact that the molecules 
in $j=1$ have generally long spontaneous emission times, on the order of $\sim 100$ s
\cite{Gonzalez-Martinez_PRA_96_032718_2017}.

We consider bosonic $^1 \Sigma^+$ alkali dipolar molecules in the vibrational state $v=0$
whith a permanent electric dipole moment $d$.
This study can be generalized to fermionic ones.
To describe the collisions between ultracold molecules,
we use a time-independent quantum formalism 
\cite{Quemener_BookChapter_2017,Wang_NJP_17_035015_2015,Gonzalez-Martinez_PRA_96_032718_2017}
including the rotational structure of the molecules 
described by a properly symmetrized and normalized
basis set $|j_1 \, m_{j_1}, j_2 \, m_{j_2}\rangle_{_\pm}$,
symmetric ($+$) 
or anti-symmetric ($-$) under permutation
of the identical molecules.
We include the rotational states $j_1=0,1$, $j_2=0,1$.
Additionally, a partial wave expansion of the collisional wavefunction
described by a spherical harmonics 
basis set $|l \, m_l\rangle$ is used.
The partial waves taken into account are $l=0,2,4$.
We numerically solve a set of close-coupled Schr{\"o}dinger equations
and by applying asymptotic boundary conditions, we
extract the scattering matrix $S$ from which we can deduce
the scattering length and the experimental observables 
such as the cross sections and the rate coefficients \cite{Quemener_BookChapter_2017}.
The complex scattering length $a$ is related to the lowest entrance channel scattering matrix element $S_{00}$ by \cite{Hutson_NJP_9_152_2007}:
\begin{eqnarray}
a = \frac{1}{i \, {k}} 
  \left( \frac{1-S_{00}({k})}{1+S_{00}({k})} \right) \bigg|_{{k} \to 0} ,
\end{eqnarray}
where $k = \sqrt{2 \mu E_c / \hbar^2}$ is the wavevector,
$E_c$ the collision energy and $\mu=m/2$ the reduced mass between the molecules 
($m$ being the mass of a molecule).
In order to reproduce available experimental data of either reactive
\cite{Ospelkaus_S_327_853_2010,Ni_N_464_1324_2010}
and non-reactive \cite{Ye_SA_4_eaaq0083_2018} molecular collisions, we impose that when two molecules come close to each other at a short distance, they are lost with a full unit probability 
\cite{Quemener_BookChapter_2017,Wang_NJP_17_035015_2015,Gonzalez-Martinez_PRA_96_032718_2017}.
To include the electromagnetic field, we employ a quantized formalism of the field,
described by a basis $|\bar{n} + n \rangle$
(see \cite{Napolitano_PRA_55_1191_1997,DeMille_EPJD_31_375_2004,Alyabyshev_PRA_80_033419_2009} for more details). 
This corresponds 
to the number of photons in the quantized field reservoir 
for a given mode $\hbar \omega$, with $|n| \ll \bar{n}$. $\bar{n}$ is a mean number (and is omitted hereafter in the notations), 
$n$ represents the number of photon lost from the quantized field and absorbed by the molecule if $n <0$
or gained by the quantized field and emitted by the molecule if $n > 0$. 
In the numerical calculation, we consider $n=0,\pm1,\pm2$. 
As for the optical shielding to take place, we consider a blue-detuned microwave 
with respect to the first rotational excited state of the molecules,
with a $\sigma^+$ circular polarization characterized by a quantum number $p=+1$ 
\cite{Napolitano_PRA_55_1191_1997,DeMille_EPJD_31_375_2004,Alyabyshev_PRA_80_033419_2009}.

As many experimental groups are now forming ultracold dipolar alkali molecules, we do not restrict our study to a specific system and rather employ a general, adimensional approach to 
treat the molecules on a same basis. The description of a molecule is
based on the combined values of $B$, $d$ and $\mu$.
To keep the study adimensional, we do not include the hyperfine structure.
The effect of the hyperfine structure has been explored in \cite{Karman_arXiv_1806_3608_2018},
where it is shown that for sufficiently high magnetic fields, the hyperfine structure can be safely neglected.
We employ the same adimensional approach 
than our previous study on shielding ultracold dipolar molecules
in an electric dc-field \cite{Gonzalez-Martinez_PRA_96_032718_2017}.
Here, the dc-field is replaced by an ac-field ${E}_{ac}$. 
We rescale the Schr{\"o}dinger equation 
using the characteristic dipolar length $s_{r_{3}} = \frac{2\mu}{\hbar^2} \frac{d^2}{4\pi\epsilon_0}$ and 
the characteristic dipolar energy $s_{E_{3}} = \frac{\hbar^2}{2 \mu \, s^2_{r_3}}$ \cite{Gao_PRL_105_263203_2010}.
The values of $s_{r_{3}}$ and $s_{E_3}$ for different alkali dipolar molecules can be found in \cite{Gonzalez-Martinez_PRA_96_032718_2017}.
We then extract four key parameters in the set of close-coupling rescaled Schr{\"o}dinger equations
very similar to the ones in our previous dc-field study.
A first parameter is a rescaled rotational constant:
\begin{eqnarray}
\tilde{B} = \frac{B}{s_{E_3}}  = \frac{8 B \mu^3}{\hbar^6} \left(\frac{d^2}{4\pi\epsilon_0}\right)^2. 
\label{tildeB}
\end{eqnarray}
Another parameter is a rescaled ac-field $\tilde{E}_{ac} = d{E}_{ac} / {B}$. 
From the usual expression of the Rabi frequency $\Omega = d{E}_{ac} / {\hbar}$, 
one can define a rescaled Rabi frequency:
\begin{eqnarray}
\tilde{\Omega} = \frac{\Omega}{B/\hbar} = \frac{d{E}_{ac}}{B} \equiv \tilde{E}_{ac} \label{tildeO}
\end{eqnarray}
which becomes the second parameter and identifies with the rescaled ac-field. 
A third parameter corresponds to a rescaled detuning:
\begin{eqnarray}
\tilde{\Delta} = \frac{\Delta}{B} = \frac{\hbar \omega - 2B}{B}. \label{tildeDelta} 
\end{eqnarray}
In this study, we 
fix this third parameter to an arbitrary positive constant of $\tilde{\Delta} = 0.025$ (blue-detuned). 
The effect of the detuning has been studied in \cite{Karman_arXiv_1806_3608_2018}.
Finally, the fourth parameter is a rescaled collision energy
$\tilde{E}_c = {E_c}/{s_{E_3}}$. 
To get rid of the collision energy dependence in our study, we consider the Wigner regime 
as $E_c \to 0$ and where the scattering length is independent of the collision energy.
The adimensional study entails a rescaled scattering length: 
\begin{eqnarray}
\tilde{a} = \tilde{a}_\mathrm{re} - i \, \tilde{a}_\mathrm{im} = \frac{a}{s_{r_3}}.
\end{eqnarray}
The ratio $\gamma$ 
of the elastic over the quenching rate coefficient 
(see Ref.\cite{Gonzalez-Martinez_PRA_96_032718_2017})
is given in term of the rescaled scattering length by:
\begin{eqnarray}
\gamma = \frac{{\beta}_\mathrm{el}}{{\beta}_\mathrm{qu}} 
= \frac{|{a}|^2}{{a}_\mathrm{im}} \, {{k}}
= \frac{|\tilde{a}|^2}{\tilde{a}_\mathrm{im}} \, {\tilde{k}}
\label{eq:ratio}
\end{eqnarray}
where ${\tilde{k}} = \sqrt{\tilde{E}_c} = \sqrt{E_c/s_{E_3}}$.

\begin{figure}[b]
 \includegraphics[width=70mm]{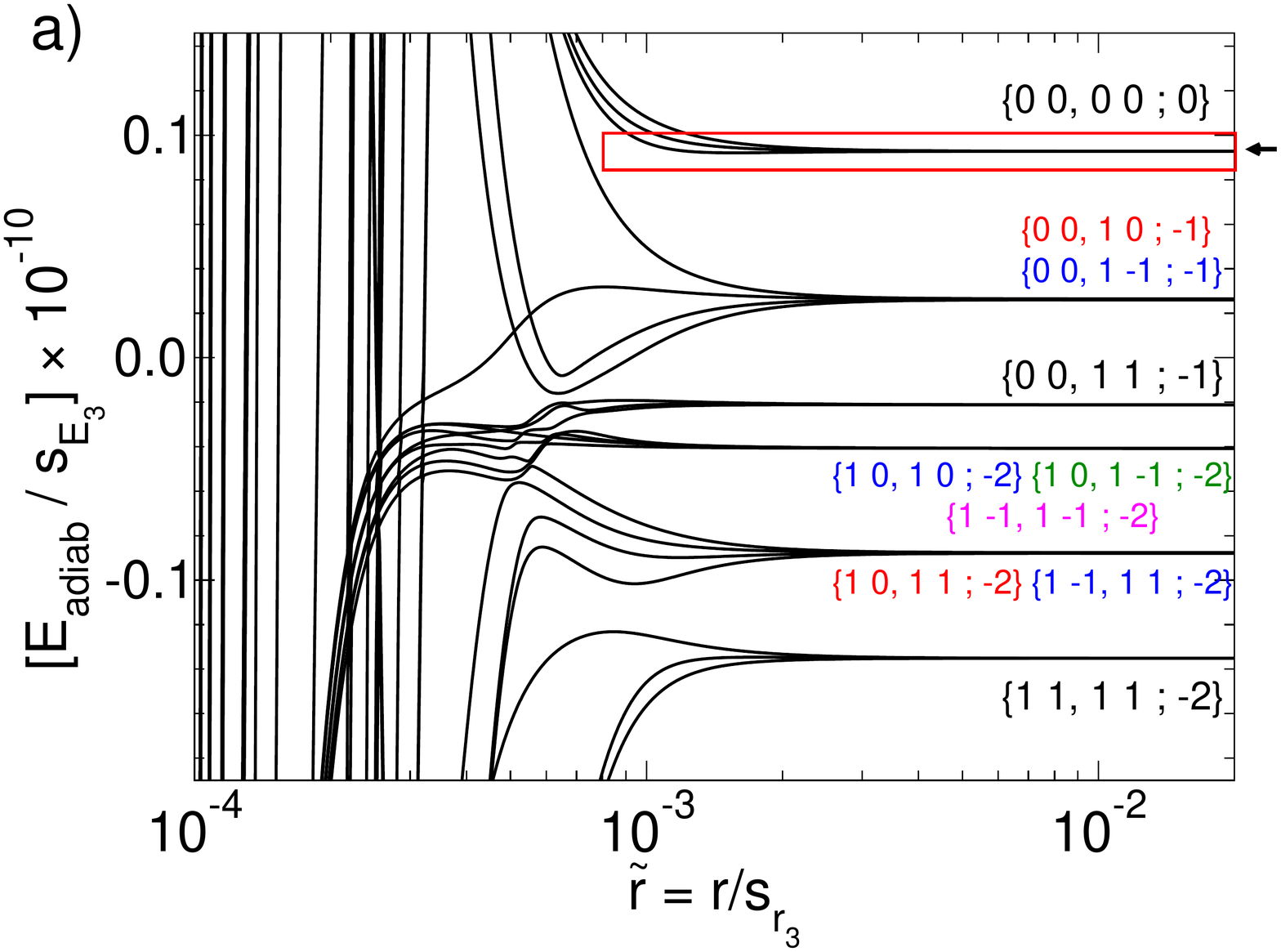} \\
 \includegraphics[width=70mm]{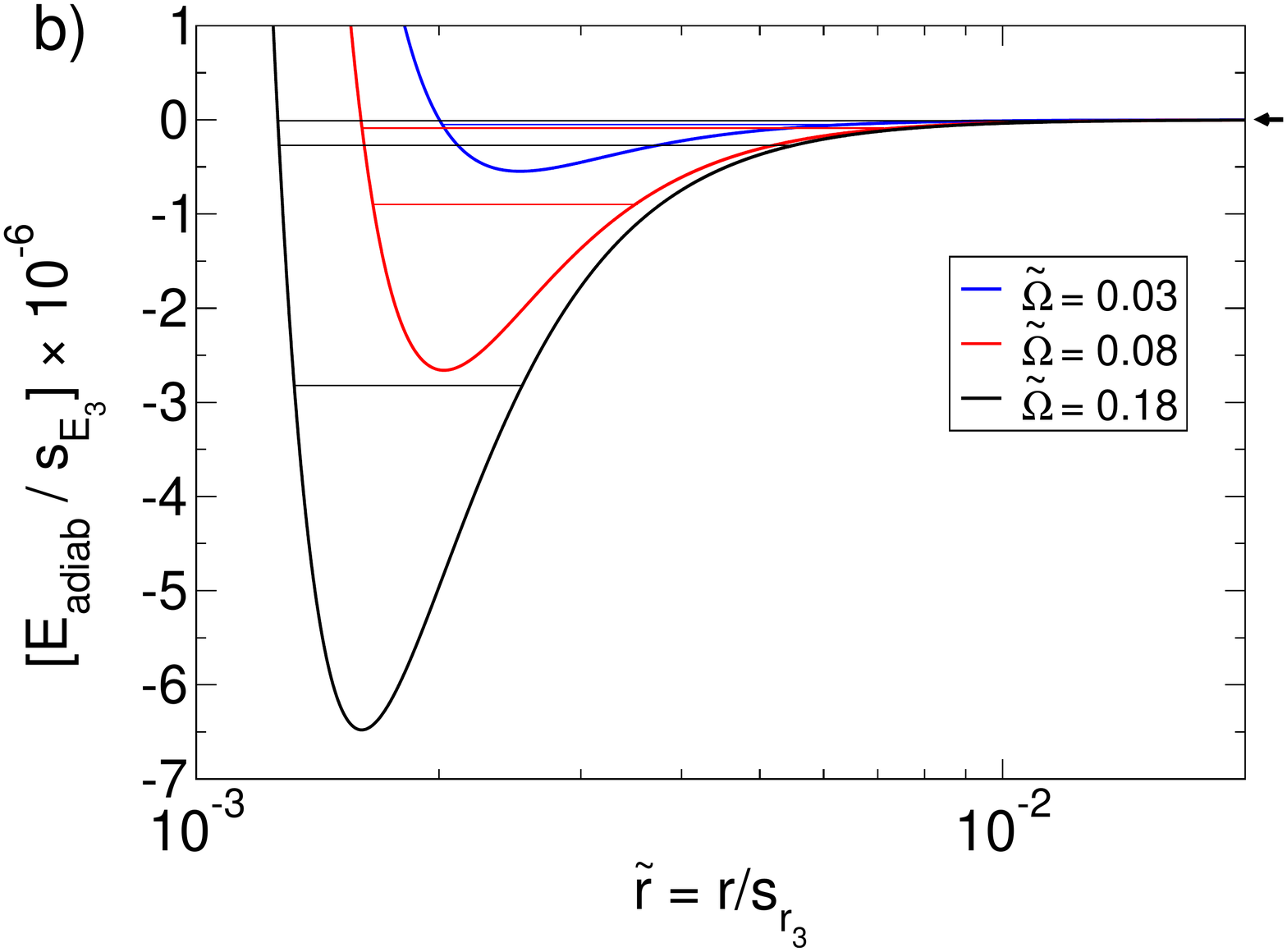}
 \caption{(Color online). 
 Top panel: Rescaled adiabatic energies as a function of the rescaled distance
 between the molecules for $\tilde{B} = 10^{10}$ ($\sim \,$NaRb), $\tilde{\Omega} = 0.18$,
 and $\sigma^+$ circularly polarized field $p=+1$. 
 The region in the red box is shown in the bottom panel. The notation  
 $\{{j}_1 \, {m}_{j_1}, {j}_2 \, {m}_{j_2} ; {n} \}$ is used to
 represent the asymptotic dressed states. The labels 
 in black (resp. red, blue, green, magenta) 
 corresponds to values of $m_\mathrm{mol_1+mol_2+field} =  m_{j_1} + m_{j_2} + n \times p = 0$ (resp. -1, -2, -3, -4)
 of the dressed system \{molecule 1 + molecule 2 + field\}. 
 Bottom panel: Close-up of the long-range potential well in the 
 lowest entrance channel for $\tilde{B} = 10^{10}$ and $\tilde{\Omega} = 0.18$ (black), $\tilde{\Omega} = 0.08$ (red), 
 $\tilde{\Omega} = 0.03$ (blue) together with the corresponding bound states energies they can support.
}
 \label{fig-spag}
\end{figure}

\begin{figure}[t]
 \includegraphics[width=90mm]{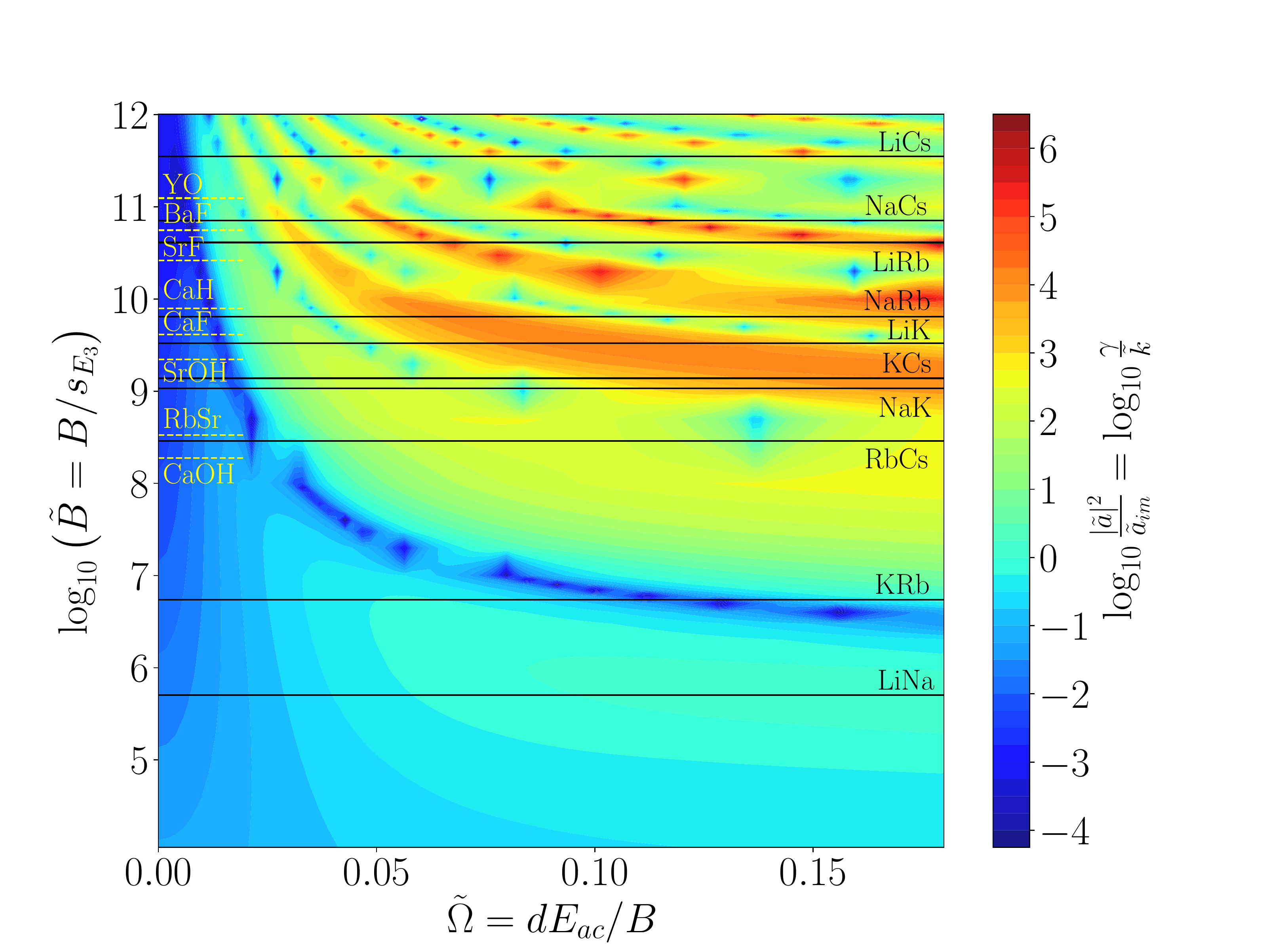}
 \caption{(Color online). $|\tilde{a}|^2/\tilde{a}_\mathrm{im} \equiv \gamma/\tilde{k}$ 
 as a function of $\tilde{B}$ and $\tilde{\Omega}$. The color scale, presented at the right of the picture, goes from $10^{-4}$ to $10^{6}$. The $\tilde{B}$ values of some characteristic dipolar molecules are reported on the figure.}
 \label{fig-gamma}
\end{figure}

We consider molecules initially prepared in their ground rotational state
$|0 0, 0 0 \rangle_{_+}$ and $|n=0\rangle$. Only the symmetric states exist 
for same, indistinguishable states and are coupled to other symmetric states.
The quantum states 
$|j_1 \, m_{j_1}, j_2 \, m_{j_2}\rangle_{_+} \, |n\rangle$
get mixed by the interaction of the molecules with the
ac-field \cite{Napolitano_PRA_55_1191_1997,DeMille_EPJD_31_375_2004,Alyabyshev_PRA_80_033419_2009}
and give rise to dressed asymptotic states,
denoted 
$\{|j_1 \, m_{j_1}, j_2 \, m_{j_2}\rangle_{_+} \, |n\rangle\}$.
This notation means that 
they tend to the undressed state
$|j_1 \, m_{j_1}, j_2 \, m_{j_2}\rangle_{_+} \, |n\rangle$ when $\tilde{\Omega} \to 0$.
They are characterized by well-defined projection numbers
$m_\mathrm{mol_1+mol_2+field} = m_{j_1} + m_{j_2} + n \times p$ (with $n,p$ being signed integer numbers)
of the dressed system \{molecule 1 + molecule 2 + field\}.
The dipole-dipole interaction will further couple the collisional states
$\{|j_1 \, m_{j_1}, j_2 \, m_{j_2}\rangle_{_+} \, |n\rangle\} \, |l \, m_l \rangle$
all together. 
The total projection number $M = m_\mathrm{mol_1+mol_2+field} + m_l$ is conserved during the collision.
For the study of the scattering length at ultra-low energies and 
given our initial state with $m_\mathrm{mol_1+mol_2+field}=0$,
we consider the lowest projection $M=0$, which implies $m_l=0$.

The dipole-dipole couplings result 
in adiabatic effective potentials illustrated in Fig.~\ref{fig-spag}-a
as a function of the rescaled distance $\tilde{r} = r/s_{r_3}$ between the molecules,
for an example at $\tilde{B} = 10^{10}$ and $\tilde{\Omega} = 0.18$. 
Strong repulsive curves arise in the initial entrance channel 
$\{|0 \, 0, 0 \, 0\rangle_{_+} \, |0\rangle\}$
indicated by an arrow, explaining more quantitatively the scheme in Fig.~\ref{fig-scheme}.
This prevents the molecules to come close to each other and being lost from chemical reactions \cite{Ospelkaus_S_327_853_2010,Ni_N_464_1324_2010} or from long-lived tetramer complexes
\cite{Ye_SA_4_eaaq0083_2018} at short-range. 
In addition, states of lower energy exist,
corresponding to the excitation of one (resp. both) of the molecules in a specific
$j, m_j$ state due to the absorption
of one (resp. two) photon lost by the quantized field with $n = -1$ (resp. $n = -2$).
When $\tilde{\Omega}$ is increased, these states
get far away from the entrance channel thus preventing inelastic 
transitions to occur.
The quenching collisions (short-range losses + inelastic processes) are expected to 
be suppressed with $\tilde{\Omega}$, explaining the mechanism of the the microwave shielding.

In Fig.~\ref{fig-gamma}, we present the quantity $|\tilde{a}|^2/\tilde{a}_\mathrm{im}$ 
which represents the ratio $\gamma$ 
when $\tilde{k} = 1$, that is at a typical collision energy of $E_c = s_{E_3}$. 
To get the ratio at $E_c > s_{E_3}$, one has to multiply this quantity by $\tilde{k}$.
For evaporative cooling techniques, $\gamma$ has to reach a factor of $~ 10^3$ or more for the process to be highly efficient. Therefore, the regions of the graph in yellow, orange and red correspond to favorable conditions for evaporative cooling. The regions in green
and blue correspond to unfavorable conditions. The rescaled Rabi frequency is plotted in abscissa and represents the amount of the ac-field applied. The rescaled rotational constant is plotted in ordinate and uniquely characterizes a molecule. The values of the dipolar alkali molecules have been reported. For indication, we also report values for $^2 \Sigma^+$ molecules of current experimental interest 
\cite{Norrgard_PRL_116_063004_2016,Truppe_NP_42_41_2017,Anderegg_PRL_119_103201_2017,Kozyryev_PRL_118_173201_2017,
Hummon_PRL_110_143001_2013,Weinstein_N_395_148_1998,Pasquiou_PRA_88_023601_2013,Chen_PRA_94_063415_2016}.
Looking at the general feature of the figure, one can distinguish two main regions for the dipolar molecules: 
a region for which $\tilde{B} > 10^8$ where the ratio can globally reach $10^3$ or more, and a region for which 
$\tilde{B} < 10^7$ where the ratio barely reach $10^2$. The former region includes the molecules RbCs, NaK, KCs, LiK, NaRb, LiRb, NaCs, LiCs and determines the good candidates for the microwave shielding. This figure also confirms the results of \cite{Karman_arXiv_1806_3608_2018} for the RbCs and KCs molecules.
The latter region includes the molecules KRb and LiNa for which the microwave shielding will be not efficient. This is due to an unfortunate combination of mass, dipole moment and rotational constant yielding a too low value of $\tilde{B}$.

\begin{figure}[t]
 \includegraphics[width=70mm]{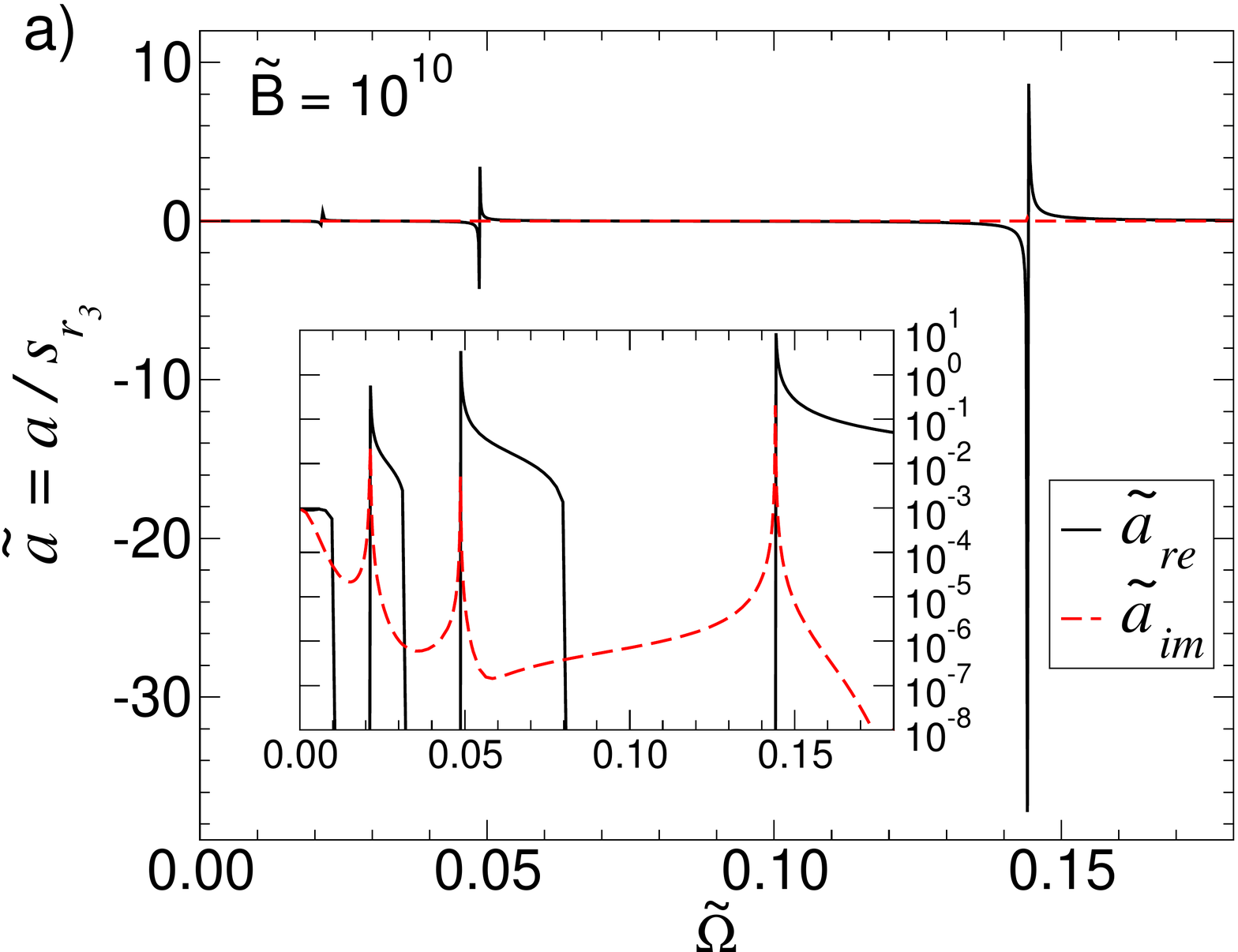} \\
 \includegraphics[width=70mm]{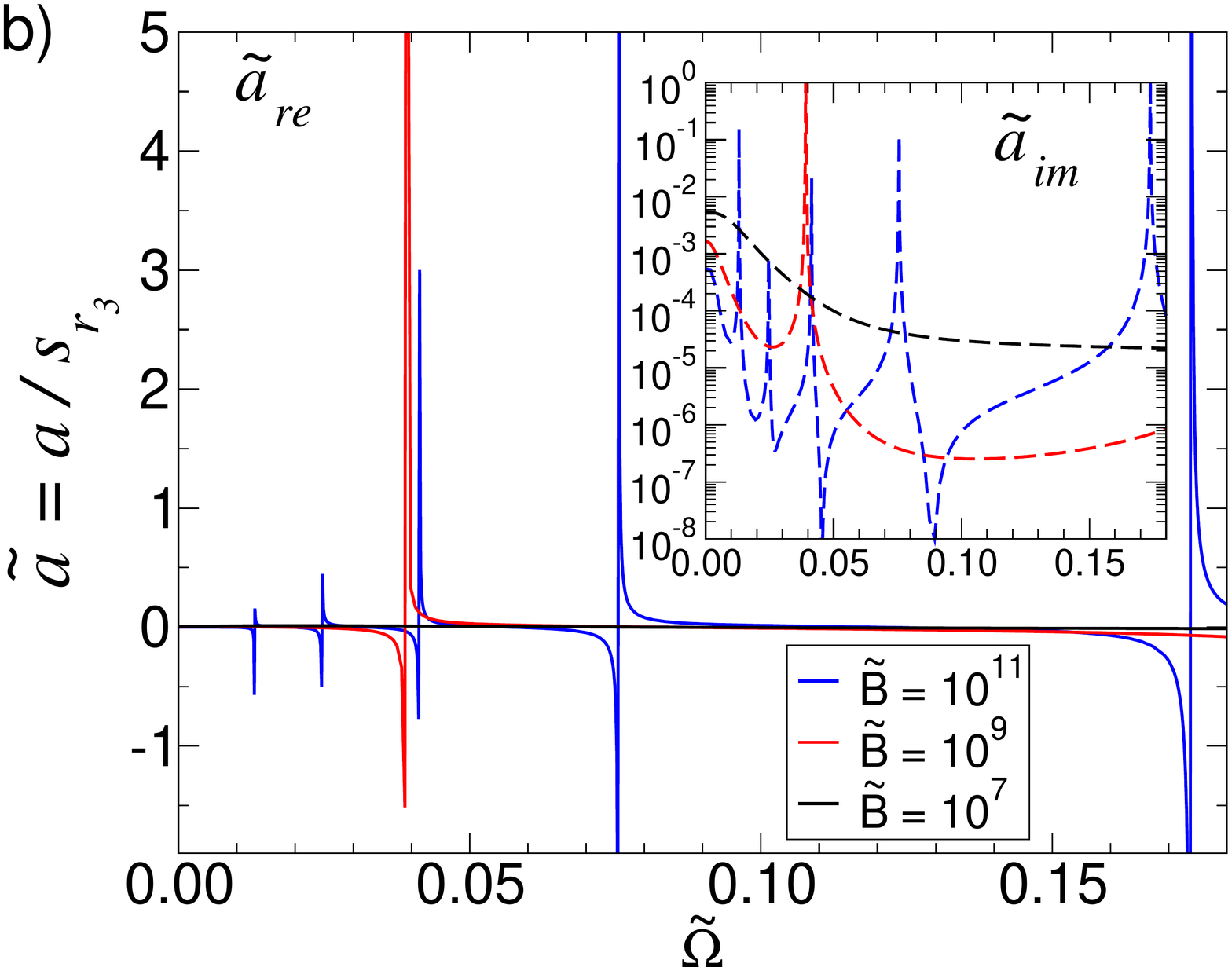}
 \caption{(Color online). Top panel: Rescaled scattering length $\tilde{a}$ 
 as a function of $\tilde{\Omega}$ for $\tilde{B}=10^{10}$ ($\sim \,$NaRb).
 Bottom panel: Same for $\tilde{B}=10^{7}$ ($\sim \,$KRb), $\tilde{B}=10^{9}$ ($\sim \,$NaK, KCs), $\tilde{B}=10^{11}$ ($\sim \,$NaCs).
 }
 \label{fig-sl}
\end{figure}

In Fig.~\ref{fig-sl}-a, we plot $\tilde{a}_{re}$ and $\tilde{a}_{im}$ as a function of $\tilde{\Omega}$
for a value $\tilde{B} = 10^{10}$ ($\sim \,$NaRb). 
There are values of $\tilde{\Omega}$, hence of the ac-field, for which the
real part $\tilde{a}_{re}$ can take large values while the imaginary part 
$\tilde{a}_{im}$ remains low (see the inset of figure).
The imaginary part globally decreases when $\tilde{\Omega}$ increases,
confirming that the quenching rate coefficients, which are proportional to
$\tilde{a}_{im}$ \cite{Gonzalez-Martinez_PRA_96_032718_2017}, also decreases as expected from
the discussion of the adiabatic curves in Fig.~\ref{fig-spag}-a.
The resonant features are explained by the apparition of a long-range, isolated shallow potential well
in the entrance channel when $\tilde{\Omega}$ is increased. 
This is illustrated in Fig.~\ref{fig-spag}-b which is a close-up of the lowest entrance channel
of Fig.~\ref{fig-spag}-a. 
At $\tilde{\Omega} = 0.18$ (black curve), the well
can support three bound states shown on the figure. 
If $\tilde{\Omega}$ is decreased, the depth of the well also decreases and those bound states 
can disappear. 
For example down at $\tilde{\Omega} = 0.08$ (red curve), the well supports now only two bound states
and at $\tilde{\Omega} = 0.03$ (blue curve), it supports only one.
When the bound states are localized at the zero energy threshold, typically for values of 
$\tilde{\Omega}$ slightly below 0.18, 0.08, 0.03, 
$\tilde{a}_{re}$ turns from a large and positive value to a large and negative value, as seen in Fig.~\ref{fig-sl}-a.

We present in Fig.~\ref{fig-sl}-b the trend of the scattering length
for increasing values of $\tilde{B}=10^7, 10^9, 10^{11}$. 
For a small value of $\tilde{B} = 10^{7}$ ($\sim \,$KRb, black curve), one cannot see any 
resonant features of $\tilde{a}$ for the present range of $\tilde{\Omega}$.
When $\tilde{B}$ is increased, typically for $\tilde{B} \ge 10^{8}$,
the long-range wells are deep enough to support bound states,
and resonant features appear in the scattering length
as in Fig.~\ref{fig-sl}-a.
This is shown for $\tilde{B} = 10^{9}$ ($\sim \,$NaK, KCs, red curve) and $\tilde{B} = 10^{11}$ ($\sim \,$NaCs, blue curve). 
These long-range bound states 
are actually reminiscent of the so-called field-linked states 
\cite{Avdeenkov_PRL_90_043006_2003,Avdeenkov_PRA_69_012710_2004}
in collisions of dipolar molecules in a static electric field.
The presence of these microwave field-linked states in the long-range wells, 
when the condition $\tilde{B} \ge 10^{8}$ is satisfied, is therefore 
responsible for the control of the scattering length value of dipolar molecules.

Technological set-ups of microwave cavities \cite{DeMille_EPJD_31_375_2004} 
are experimentally tractable nowadays \cite{Dunseith_JPBAMOP_48_045001_2015}.
Input powers of the order of $\sim \,$kW yield corresponding ac-fields of $\sim \,$10~kV/cm. 
This is already highly sufficient to what is needed for alkali dipolar molecules with $\tilde{B} \ge 10^8$
presented in the study. For example, at $\tilde{\Omega}=dE_{ac}/B=0.18$, one needs
at most $E_{ac} \sim \,$1 kV/cm for the lowest value of $d/B$ (LiK molecule).
The microwave energies available are in the range [2 - 18 GHz] which correspond 
exactly to the energies needed (twice the rotational constant of the alkali molecules).
Finally, better control over circular polarization fields becomes 
nowadays possible \cite{Signoles_PRL_118_253603_2017}.
Therefore, with the current improvement of the microwave technologies,
the control of the scattering length of 
dipolar molecules seems experimentally realistic, 
and will certainly open a new regime of 
strongly interacting and correlated physics with ultracold dipolar molecules.

\begin{acknowledgments}
We acknowledge fundings from the FEW2MANY-SHIELD project \#~ANR-17-CE30-0015,
the COPOMOL project \#~ANR-13-IS04-0004 and the 
BLUESHIELD project \#~ANR-14-CE34-0006 from Agence Nationale de la Recherche.
We also acknowledge fruitful and stimulating discussions with the Th{\'e}omol team members,
especially M. L. Gonz{\'a}lez-Mart\'{\i}nez, A. Orb{\'a}n, M. Lepers, O. Dulieu and N. Bouloufa-Maafa.
\end{acknowledgments}

\bibliography{../../../BIBLIOGRAPHY/bibliography.bib}

\end{document}